\documentclass[%
reprint,prx,
superscriptaddress,
 amsmath,amssymb,
 aps,
floatfix,
]{revtex4-1}
\usepackage{nicefrac}
\usepackage{graphicx,times}
\usepackage{dcolumn}
\usepackage{bm}
\usepackage{hyperref}
\usepackage{xcolor}
\usepackage[mathlines]{lineno}

\expandafter\ifx\csname package@font\endcsname\relax\else
 \expandafter\expandafter
 \expandafter\usepackage
 \expandafter\expandafter
 \expandafter{\csname package@font\endcsname}%
\fi

\def\be{\begin{equation}}
\def\ee{\end{equation}}
\def\bea{\begin{eqnarray}}
\def\eea{\end{eqnarray}}
\def\bsplit{\begin{split}}
\def\esplit{\end{split}}

\begin{document}
\raggedbottom

\title{Interplay between substrate rigidity and tissue fluidity regulates cell monolayer spreading}
\author{Michael F. Staddon}
\affiliation{Center for Systems Biology Dresden, Dresden, Germany}
\affiliation{Max Planck Institute of Molecular Cell Biology and Genetics, Dresden, Germany}
\affiliation{Max Planck Institute for the Physics of Complex Systems, Dresden, Germany}
\author{Michael P. Murrell}
\affiliation{Department of Biomedical Engineering and Department of Physics, Yale University, New Haven, CT, USA}
\author{Shiladitya Banerjee}
\email{shiladtb@andrew.cmu.edu}
\affiliation{Department of Physics, Carnegie Mellon University, Pittsburgh, PA, USA}


\begin{abstract}
\noindent Coordinated and cooperative motion of cells is essential for embryonic development, tissue morphogenesis, wound healing and cancer invasion. A predictive understanding of the emergent mechanical behaviors in collective cell motion is challenging due to the complex interplay between cell-cell interactions, cell-matrix adhesions and active cell behaviors. To overcome this challenge, we develop a predictive cellular vertex model that can delineate the relative roles of substrate rigidity, tissue mechanics and active cell properties on the movement of cell collectives. We apply the model to the specific case of collective motion in cell aggregates as they spread into a two-dimensional cell monolayer adherent to a soft elastic matrix. Consistent with recent experiments, we find that substrate stiffness regulates the driving forces for the spreading of cellular monolayer, which can be pressure-driven or crawling-based depending on substrate rigidity. On soft substrates, cell monolayer spreading is driven by an active pressure due to the influx of cells coming from the aggregate, whereas on stiff substrates, cell spreading is driven primarily by active crawling forces. Our model predicts that cooperation of cell crawling and tissue pressure drives faster spreading, while the spreading rate is sensitive to the mechanical properties of the tissue. We find that solid tissues spread faster on stiff substrates, with spreading rate increasing with tissue tension. By contrast, the spreading of fluid tissues is independent of substrate stiffness and is slower than solid tissues. We compare our theoretical results with experimental results on traction force generation and spreading kinetics of cell monolayers, and provide new predictions on the role of tissue fluidity and substrate rigidity on collective cell motion.

\end{abstract}

\maketitle

\section{Introduction}

\noindent Tissue spreading is a fundamental biological process underlying collective cell movement during development~\cite{friedl2009collective,behrndt2012forces,julicher2017emergence,maniou2021hindbrain}, cancer invasion~\cite{leber2009molecular,carey2012mechanobiology,friedl2012classifying,labernadie2017mechanically}, and wound healing~\cite{martin2004parallels,brugues2014forces,ajeti2019wound,tetley2019tissue}. The collective motion of cells during tissue spreading is regulated by the interplay between cell-cell and cell-matrix adhesions~\cite{collins2015,friedl2017,de2017}, as well as by active processes such as lamellipodial cell crawling and actomyosin contractility that control the dynamic mechanical properties of individual cells and tissues~\cite{salbreux2012actin,ladoux2017mechanobiology}. Many of the active mechanical components of cells are mechanosensitive and interact with each other via complex feedback networks~\cite{huveneers2013,case2015,iyer2019,cavanaugh2020,banerjee2020}, making it experimentally challenging to decipher the key regulators of cellular mechanical behaviors~\cite{discher2005tissue,ladoux2016front}.

A common model system for studying the mechanics of collective cell migration is the spreading of a three-dimensional cell aggregate over a soft elastic substrate~\cite{douezan2011spreading,beaune2014cells,beaune2017reentrant,perez2019active,yousafzai2020tissue,yousafzai2022active}. When placed onto an adhesive substrate, the aggregate spreads out in a process similar to the wetting of liquid droplets, in which differences in adhesion between cell-cell and cell-substrate contacts drives the spreading of the fluid aggregate~\cite{ryan2001tissue,gonzalez2012soft,beaune2014cells,beaune2017reentrant}. However, the spreading dynamics of a living tissue is more complex than the wetting of passive liquid droplets. Recent work has demonstrated the importance of cellular mechanics and intercellular adhesions in regulating the spreading dynamics of tissues, whose mechanical properties can range from fluids to glassy jammed solids~\cite{trepat2009physical,angelini2011glass,park2015unjamming,tetley2019tissue}. Both tissue viscosity and cell-cell adhesion strengths are regulated by E-cadherins~\cite{van2008,lecuit2015,iyer2019}. Upon reduction in E-cadherin expression, the spreading rate of the cell aggregate is elevated~\cite{douezan2011spreading}, while increasing E-cadherin expression~\cite{perez2019active} or substrate stiffness can induce dewetting of already spread aggregates~\cite{beaune2018}.

Active, non-equilibrium behavior of cells is another key regulator of collective cell spreading not accounted for in the wetting model and remains poorly understood. However, recent experiments have begun to uncover how cellular aggregates adapt to the mechanics of the extracellular matrix in order to drive robust collective motion~\cite{yousafzai2020tissue}. Depending on matrix rigidity, cells may polarize, generating active traction stresses to crawl outwards. The stiffness of the substrate is important not only for providing passive friction to cell motion, but can also induce cell polarization~\cite{ladoux2016front,gupta2019cell}. A recent study showed that cellular aggregates can tune their mechanics and migratory behaviors depending on matrix rigidity~\cite{yousafzai2020tissue}. On stiff substrates, traction stresses are elevated at the tissue boundary, driving rapid outward motion of cells. By contrast, on soft substrates, traction stresses are attenuated and cell spreading in driven by an outward active pressure.

In this paper, we develop a cell-based active vertex model for cellular monolayers to investigate the role of substrate stiffness and tissue mechanics in monolayer spreading. A variety of theoretical models and methods have been developed in recent years to describe the collective motility of cells, including continuum models~\cite{kopf2013,banerjee2015,notbohm2016,blanch2017,alert2018,banerjee2019}, particle-based models~\cite{camley2016,zimmermann2016}, lattice-based models~\cite{szabo2010,swat2012}, as well as vertex-based~\cite{fletcher2014,barton2017,staddon2018} and voronoi models~\cite{bi2016}. While continuum models of tissue spreading have been successful in predicting traction force organization~\cite{banerjee2019}, wave-like dynamics~\cite{banerjee2015} and wetting transitions~\cite{douezan2011spreading,alert2018,perez2019active}, these models assume fixed constitutive relations for tissue materials properties, and thus do not account for dynamic changes in tissue mechanical properties due to single-cell level active mechanical behaviors. On the other hand, discrete cell-based models for tissues have not yet been implemented to study how cellular aggregates adapt to substrate mechanical properties in order to drive collective cell motion. We bridge this gap by developing an active vertex model for cell aggregate spreading that allow us to study the interplay between tissue mechanics, substrate mechanics, as well as the role of active single-cell behaviors in the collective motility of cell monolayers. 

\begin{figure}[t]
\includegraphics[width=\columnwidth]{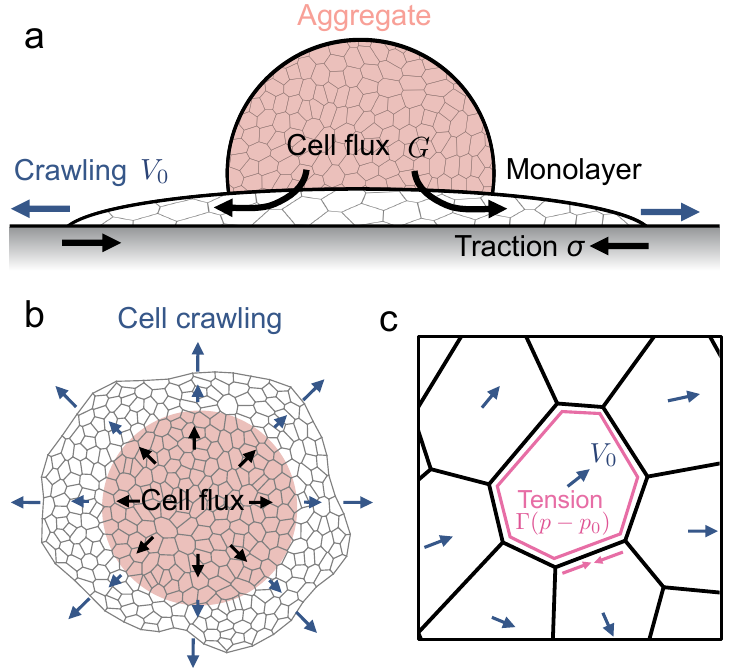}
\caption{Mechanical forces driving the spreading of a multicellular aggregate. (a) Schematic showing the side view of an aggregate spreading as a monolayer over a soft adhesive substrate. Spreading is driven by the influx of cells coming from the aggregate into the center of the monolayer, as well as by active cell crawling that generate traction stresses on the substrate. (b) Schematic showing the top view of the spreading monolayer that is modeled using an active vertex model. (c) Schematic for an active vertex model of a spreading monolayer. Blue arrows indicate the orientation of cell polarity, which indicates the direction of cell crawling. Cell edges are under tension due to actomyosin contractility.}
\label{fig:1}     
\end{figure}

We model the collective motion of cells in a three-dimensional aggregate as they spread into a two-dimensional monolayer adherent to a flat substrate (Fig.~\ref{fig:1}a). The aggregate acts as a reservoir of cells, feeding cells into the spreading monolayer by a process called permeation~\cite{beaune2014cells} (Fig.~\ref{fig:1}a). By developing an active vertex model for the monolayer, we study how changes in substrate stiffness affects the speed and the modes of monolayer spreading. Our model successfully captures the experimental behaviour initially reported by Ref.~\cite{yousafzai2020tissue}. We find that monolayer spreading on soft substrates is driven by an active pressure due to the influx of cells coming from the aggregate. By contrast, on stiff substrates, monolayer spreading is driven by active crawling of cells that generate elevated traction forces at the monolayer boundary. To conceptually understand the role of substrate stiffness in collective cell motion, we develop a simple mechanical model of the monolayer as a viscoelastic material that spreads by active crawling as well as due to growth in the preferred area of the material. This model captures the key results from the vertex model, and predicts that tissue stiffness governs the long term spreading rate of the tissue. To test this prediction, we vary both monolayer tension and substrate stiffness in the vertex model, and find that tissues with increased tension display increased sensitivity to substrate stiffness. Moreover, we find that solid-like tissues spread faster than fluid-like tissues due to a reduced bulk modulus of the tissue overall. These results provide a theoretical understanding for the role of both tissue fluidity and substrate rigidity, and their interplay, on the spreading dynamics of multicellular aggregates. 

\section{Active vertex model for collective cell spreading}
\noindent To describe the dynamics of cell monolayer spreading, we model the cell monolayer underneath the aggregate as a two-dimensional tissue using the framework of the vertex model~\cite{nagai2001dynamic,farhadifar2007influence,staple2010mechanics,alt2017vertex}. The aggregate generates a continuous flux of cells into the monolayer uniformly within a central region that marks the contact area of the aggregate (Fig.\ref{fig:1}a). Cells in the monolayer are able to actively crawl out into the free space during spreading, while the cell influx from the aggregate, modeled by stochastic insertion of new cells, increases the number of cells in the monolayer over time (Fig.\ref{fig:1}a-b). Each cell in the monolayer is modeled by a two-dimensional polygon, with edges representing the cell-cell junctions and the vertices representing tri-cellular junctions (Fig.\ref{fig:1}b-c). The mechanical energy $E$ of the monolayer is given by
\begin{equation}
E_\text{mech} = \sum_{\alpha} \frac{1}{2} K (A_\alpha - A_0)^2 + \sum_{\alpha} \frac{1}{2}\Gamma(P_\alpha- P_0)^2\;,
\end{equation}
where $\alpha$ indicates the cell number, and $A_\alpha$ and $P_\alpha$ are the area and the perimeter of the $\alpha^{th}$ cell, respectively. The first term represents the area elasticity of the cell, describing in-plane compressibility with elastic modulus $K$ and preferred area $A_0$. The second term represents a balance between cytoskeletal contractility, interfacial tension, and cell-cell adhesion, where $\Gamma$ is the contractility of the cell and $P_0$ is the preferred perimeter defined as $P_0=-\gamma/2\Gamma$, with $\gamma$ the interfacial tension on cell edges~\cite{staple2010mechanics}. Each cell contributes a resultant tension on each edge equal to $\Gamma (P_\alpha - P_0)$. The mechanical force acting on each vertex is given by the energy gradient
\begin{equation}
\mathbf{F}_i = -\frac{\partial E_\text{mech}}{\mathbf\partial \mathbf{x}_i}
\end{equation}
where $i$ indicates the vertex number, and $\mathbf{x}_i$ is the vertex position. In simulations, we non-dimensionalize force scales by $KA_0^\frac{3}{2}$ and length scales by $A_0^\frac{1}{2}$ so that our non-dimensional energy $\tilde{E}$ becomes 
\begin{equation}\label{eq:energy}
\tilde{E} = \sum_{\alpha} \frac{1}{2} (a_\alpha - 1)^2 + \sum_{\alpha} \frac{1}{2}\tilde{\Gamma}(p_\alpha - p_0)^2\;,
\end{equation}
where $a_\alpha=A_\alpha/A_0$, $p_\alpha=P_\alpha/A_0^{1/2}$, $p_0=P_0/\sqrt{A_0}$, and $\tilde{\Gamma}=\Gamma/KA_0$.

To model active cell crawling, each cell is assigned a unit polarity vector $\mathbf{p}_\alpha$ that defines the direction of cell crawling or self-propulsion. {\color{black}Self-propulsion models for cell motility have been extensively studied in the active matter literature, including in particle-based models~\cite{szabo2006,chate2008,henkes2011}, active gel models~\cite{banerjee2015,notbohm2016,peyret2019}, vertex models~\cite{barton2017,sussman2017,staddon2018,schaumann2018}, and self-propelled voronoi models for cell layers~\cite{bi2016}, with various different rules for cell polarity dynamics including random rotation~\cite{bi2016,sussman2017}, alignment with the polarity of cell neighbors~\cite{chate2008,staddon2018,schaumann2018,barton2017}, cell migration direction~\cite{szabo2006,henkes2011,barton2017} or the total force acting on each cell~\cite{peyret2019}. Here we assume that} cells on the boundary attempt to crawl into free space, setting the polarity vector to be a unit vector perpendicular to their free edge pointing outwards. Cells within the monolayer then align their polarity vector with their neighbours
\begin{equation}
\frac{{\rm d} \mathbf{p}_\alpha}{{\rm d} t} = k_p \sum_\beta \left( \frac{\mathbf{p}_\beta}{n_\beta} - \frac{\mathbf{p}_\alpha}{n_\alpha} \right)
\end{equation}
where $\beta$ labels the neighbouring cells of the cell $\alpha$, $k_p$ is the polarity alignment rate, $n_\alpha$ and $n_\beta$ are the number of neighbors for cells $\alpha$ and $\beta$ respectively. The resultant effect is a diffusive relaxation of polarity from the edge to the center of the monolayer, with cells near the center of the monolayer crawling outwards but slower, {\color{black}consistent with experimental observations of cell velocity distribution in spreading aggregates~\cite{yousafzai2020tissue}}. {\color{black} We expect using a model where polarity aligns with cell velocity~\cite{peyret2019} would give similar results, as motion is mostly radially outwards due to outward crawling and pressure from the center, and so polarity and motion are correlated.} Assuming over-damped motion, force balance at cell vertices is then given by
\begin{equation}
\mu \frac{{\rm d}\mathbf{x}_i}{{\rm d}t} = \mu v_0 \langle \mathbf{p}_\alpha \rangle_i + \mathbf{F}_i
\end{equation}
where $\mu$ is the friction coefficient, $v_0$ is the cell crawling speed, and $\langle \mathbf{p}_\alpha \rangle_i$ is the average polarity vector for cells containing vertex $i$. The traction force generated on the substrate is then given by $\mu(\frac{{\rm d}\mathbf{x}_i}{{\rm d}t} -v_0 \langle \mathbf{p}_\alpha \rangle_i) = \mathbf{F}_i$. 

{\color{black}For computational efficiency, we do not explicitly model the substrate but assume an effective model} that the cell-substrate friction $\mu$ increases monotonically with substrate elastic modulus $E$, such that $\mu = \mu_0 E$, where $\mu_0$ is a constant. {\color{black}This result follows from our previous work~\cite{staddon2018,ajeti2019wound} where we had shown that cell-substrate friction increases with substrate stiffness by explicitly modeling the substrate as an elastic network adherent to a migrating cell collective. Furthermore, the linear relationship between friction and substrate stiffness also} follows from a kinetic model of focal adhesion complexes~\cite{walcott2010}, which predicts that cell-substrate adhesive interactions provide a frictional drag that increases with the elastic modulus of the substrate. {\color{black} At high elastic modulus, the friction is likely to saturate}. Cell polarity and migration speeds are also substrate stiffness-dependent. Experimentally it was observed that actin stress fiber organization and directionality of cell movement is substrate stiffness dependent, such that cell polarization increases monotonically with substrate elastic modulus~\cite{discher2005tissue,zemel2010,trichet2012,oakes2014,yousafzai2020tissue}. While single-cell migration speeds show biphasic stiffness dependence~\cite{pathak2012,bangasser2017}, the speed of collective cell migration increases monotonically with substrate stiffness~\cite{ng2012}. Based on these observations, we assume for simplicity that cell crawling speed in a tissue increases linearly with substrate elastic modulus, $v_0 = c_0 E$, where $c_0$ is a proportionality constant, {\color{black} although typically the crawl speed would saturate at high rigidity}.
 
The influx of cells from the aggregate into the monolayer is modelled by stochastically inserting new cells into the the 2D monolayer, within a region that the aggregate occupies above the monolayer~\cite{yousafzai2020tissue} (Fig.~\ref{fig:1}). When a new cell is added, an existing cell within the aggregate area is randomly selected and subdivided into two, with the new cell initialized to have zero polarity. The cell influx, or rate of cell additions, increases with monolayer area, as the aggregate has more contact area to add cells in, until the monolayer spreads outside of the aggregate area. This results in a time-dependent cell flux $J(t)= G \min(A(t), A_\text{ag}) / a^*$, where $G$ is the maximum flux, $A(t)$ is the area of the monolayer at time $t$, $A_\text{ag}$ is the contact area of the aggregate {\color{black}at $90^{\circ}$ contact angle}, and $a^*$ is the area of a single cell at the mechanical equilibrium. Thus, without cell crawling, the monolayer area will grow exponentially until it is larger than the aggregate area, after which it will grow linearly. {\color{black} Without the addition of cells from the aggregate, the monolayer would exhibit Kelvin-Voigt type viscoelasticity, in which the aggregate would increase in area up to a maximum value. However, at longer timescales cells in the monolayer may adapt their shape to stretch and get thinner, or divide, giving a more fluid like behaviour of the tissue which is often used to model monolayers.}
\begin{figure}[t]
\includegraphics[width=\columnwidth]{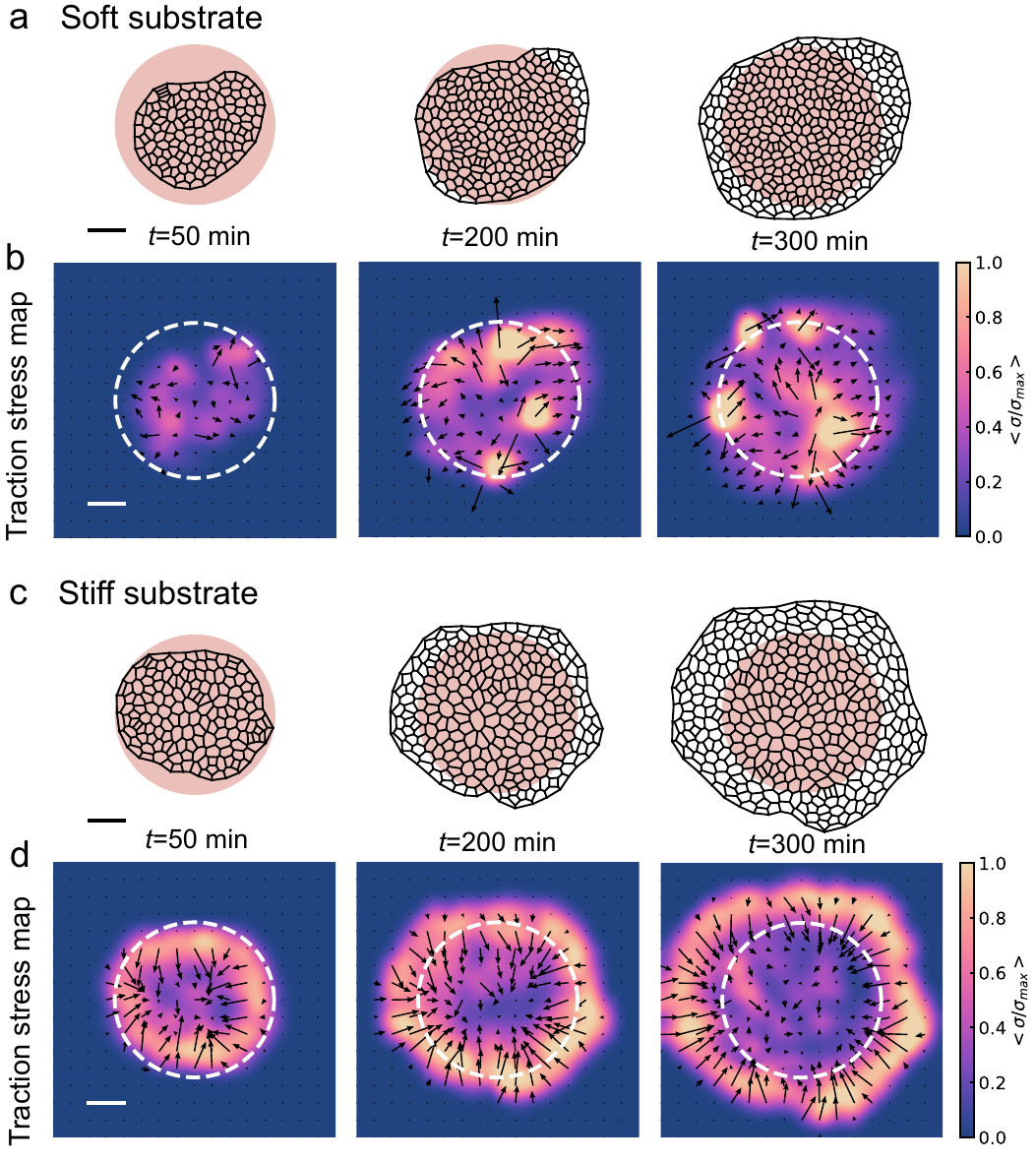}
	\caption{Cell spreading dynamics and collective migration modes are substrate rigidity-dependent. (a) Cell outlines, and (b) traction stress maps of the cell monolayer during aggregate spreading on a soft gel ($E = 5$, dimensionless units), at $t = 50$, $200$, and $350$ min. The pink shaded region indicates the contact area of the aggregate above the monolayer, at $90^{\circ}$ contact angle. See Movies 1 and 2 for a time-lapse video. (c) Cell outlines, and (d) traction stress maps of cell monolayer spreading on a stiff gel ($E = 30$), at $t = 50$, $200$, and $350$ min. See Movies 3 and 4 for a time-lapse video. Scale bar represents 5 units of length.}
	\label{fig:2}     
\end{figure}

The model is implemented in Surface Evolver~\cite{brakke1992surface} and solved numerically using the forward Euler method, with a timestep $dt = 0.05$ min. At each time step, we update the vertex position and then perform $T_1$ transitions, or neighbour swaps, for edges that become lower than a length threshold $L^* = 0.05$. To implement cell influx from the aggregate we calculate the rate of cell additions, $J(t)= G \min(A(t), A_\text{ag}) / a^*$, where $a^*$ is the mean area of a single cell at equilibrium. Then, with probability $J dt$, a random cell from the monolayer within the aggregate area is subdivided into two cells by splitting the cell with a new edge. The tissue mechanical parameters are taken from a previous study on MDCK monolayers~\cite{ajeti2019wound}, while the cell flux rate and crawling speeds are chosen to reproduce experimentally measured spreading rates of the monolayer~\cite{yousafzai2020tissue} (Table~\ref{table1}). To simulate {\color{black}monolayer} spreading, we begin with a monolayer of half the {\color{black}maximum contact} area of the aggregate, located at the center of the aggregate and beginning from a state of mechanical equilibrium. {\color{black} The aggregate remains fixed in position as the monolayer spreads out, injecting cells into the monolayer over a maximum area $A_{ag}$~\cite{yousafzai2022active}.} We then simulate cell spreading including cell flux and crawling for 400 minutes, tracking the monolayer area and traction forces at each cell vertex of the tissue.
\begin{table}[ht]
\centering
\begin{tabular*}{0.48\textwidth}{@{\extracolsep{\fill}}lll}
\hline\noalign{\smallskip}
Parameter & Symbol & Value  \\
\noalign{\smallskip}\hline\noalign{\smallskip}
Contractility & $\tilde{\Gamma}$ & 0.166 \\
Shape index & $p_0$ & 1.5 \\
Crawl speed coefficient & $c_0 = v_0 / E$ & 1 \\
Friction coefficient & $\mu_0 = \mu / E$ & 0.1 min \\
Polarity alignment rate & $k_p$ & 1 min \\
Aggregate flux rate & $G$ & 22.5 \%/ hour \\
Aggregate area & $A_\text{ag}$ & 100 \\
\noalign{\smallskip}\hline
\end{tabular*}
\caption{List of default parameter values in the vertex model.}
\label{table1}
\end{table}

\section{Substrate rigidity regulates the driving forces for collective spreading}\label{sec:1}
\noindent To understand how substrate rigidity regulates collective cell spreading, we simulated the active vertex model of a spreading monolayer at different values of substrate stiffness, $E$. Our simulations reveal two distinct mechanisms of cell monolayer spreading which are dependent on the substrate stiffness. On soft substrates, cell spreading is driven by an active pressure arising from the influx of cells from the aggregate into the center of the monolayer (Fig. \ref{fig:2}a,b, Movie 1-2). As cells are added into the monolayer, pressure builds up from the newly incorporated compressed cells, producing radially outward traction stresses localized around the center of the monolayer that increase in magnitude over time (Fig. \ref{fig:2}b, Movie 2). By contrast, monolayer spreading on stiff substrates is driven primarily by cell crawling. This results in faster spreading rates on stiff substrates as compared to soft (Fig. \ref{fig:2}c, Movie 3), and radially inward traction stresses accumulate on the border of the monolayer (Fig. \ref{fig:2}d, Movie 4). While peripherally localized traction stresses have been observed in experiments of spreading tissues before~\cite{trepat2009physical,notbohm2016,perez2019active}, and reproduced in theoretical models~\cite{banerjee2015,notbohm2016,blanch2017,perez2019active}, pressure-driven traction stresses at the center of the tissue have been recently reported in experiments on cell aggregates spreading on soft substrates~\cite{yousafzai2020tissue}.
\begin{figure}[t]
\includegraphics[width=\columnwidth]{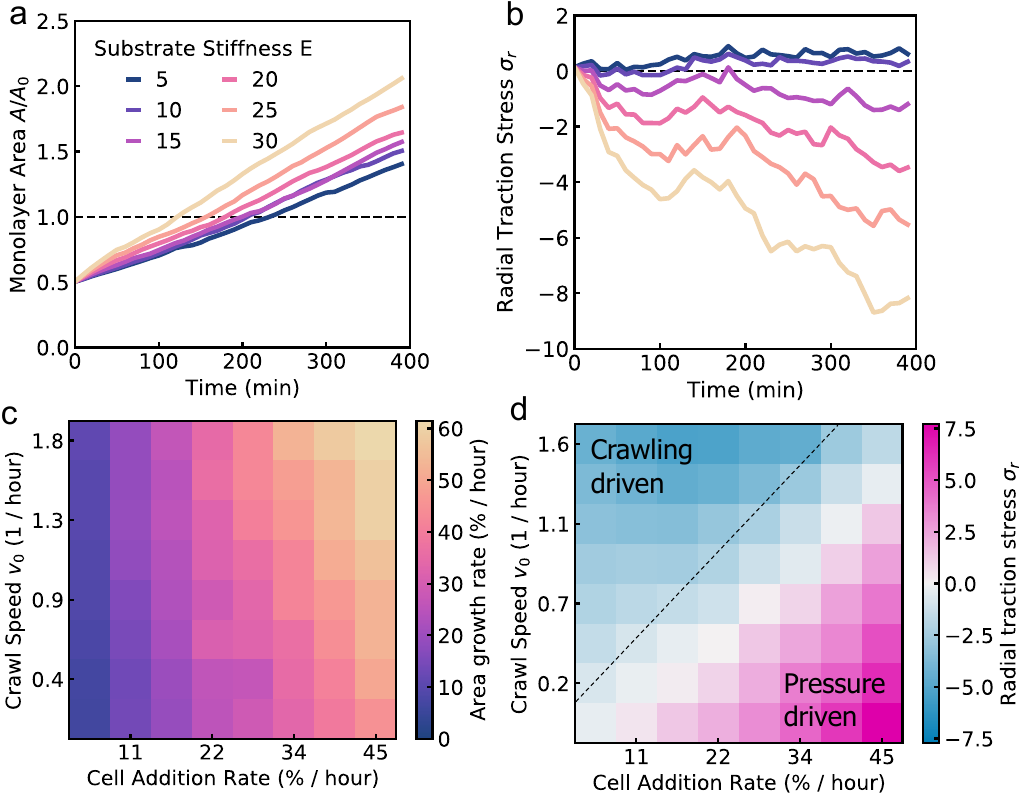}
	\caption{Cell monolayer spreading may be pressure-driven or crawling-based depending on the rigidity of the substrate. (a) Monolayer area, and (b) total radial traction stress over time for different values of substrate stiffness. (c) Heatmap of monolayer area growth rate, (d) and total radial traction stress for varying cell crawl speed and cell addition rate at an intermediate value of substrate stiffness ($E = 15$).}
	\label{fig:3}     
\end{figure}

The dynamics of monolayer area depend on the mechanism driving monolayer spreading. When cell monolayer spreading is driven by the influx of cells from the aggregate, the monolayer area initially increases exponentially, as the cell influx rate is proportional to the monolayer area (Fig.~\ref{fig:3}a). Once the monolayer spread area is larger than the contact area of the aggregate, cells are added at a constant rate, resulting in a constant rate of monolayer area increase. On a soft substrate, the radial traction stress ($\sigma_r$) builds up to positive values due to the accumulation of compressed cells near the center of the monolayer, before plateauing (Fig.~\ref{fig:3}b). As substrate stiffness increases, there is a transition to crawling-driven spreading resulting in negative (inward) traction stresses. At high substrate stiffness, the monolayer area initially increases rapidly due to border crawling forces until the cells are strained, creating large inward traction stresses (Fig.~\ref{fig:3}a). The monolayer area then increases linearly in time, at a faster rate than on soft gels, with a gradual increase in the magnitude of traction stresses (Fig.~\ref{fig:3}b).

Next, we investigate how the speed of cell crawling and the rate of cell additions control the spreading rate and the pattern of traction forces generated by the cell monolayer. At high cell addition rate, cell motion is pressure-driven (outward traction stresses), whereas at high crawling speeds, cell motion is crawling-driven (inward traction stresses). The interplay between pressure-driven and crawling-driven motion is regulated by substrate stiffness, as discussed previously (Fig.~\ref{fig:3}a-b). At a fixed (intermediate) value of substrate stiffness, we find that the area growth rate, measured by the rate of area increase as the monolayer doubles in size, increases with crawl speed, but is limited by the rate of cell addition (Fig. \ref{fig:3}c). When the crawling speed is high but the cell addition rate is low, the monolayer is unable to effectively spread. When we compute the traction forces, we find large outward tractions when the cell addition rate is high (compared to cell crawling speeds), indicating pressure-driven spreading. In the limit where cells are unable to crawl but the cell addition rate is high, spreading rates still remain high, with crawling forces providing a small boost to the overall spreading rate while reducing the magnitude of traction stresses (Fig. \ref{fig:3}d). By contrast, large inward traction forces are generated as cells stretch during spreading driven primarily by crawling forces (Fig. \ref{fig:3}d). In this case, spreading rates are limited by the lower rate of stress relaxation in the form of newly added cells.

\section{Continuum model for collective cell spreading reveals the relative roles of substrate rigidity and tissue mechanics}
\label{sec:2}
\noindent While the vertex model simulations describe the role of substrate rigidity and cell influx on traction force generation and monolayer spreading rate, it does not immediately reveal an intuitive understanding of the relative roles of substrate rigidity and tissue mechanics on cell monolayer spreading. To this end, we develop a continuum mean-field model for the spreading of an elastic monolayer, neglecting spatial variations in dynamics for simplicity. We consider a monolayer of $N(t)$ cells of area $a_i(t)$ each, where $1\leq i\leq N(t)$. Each cell is self-propelled along their polarity vector at a speed $v_0$, and $K$ is the area compressibility modulus or a 2D bulk modulus of each cell. Total mechanical energy of the cells in the monolayer is given by
\begin{equation}
E_\text{mech}=\sum_{i=1}^{N(t)} \frac{K}{2}(a_i(t)-a_0)^2
\end{equation}
where $a_0$ is the target area of each cell. During migration, mechanical force in each cell is balanced by active and dissipative forces. In the linear response regime, the rate of energy dissipated by monolayer spreading is given by
\begin{equation}
\mathcal{D}=\sum_{i=1}^{N(t)} \frac{\mu}{2h^2}\dot{a}_i(t)^2\;,
\end{equation}
where $\mu$ is the friction between the cell and the substrate, and $h$ is the average height of the cell monolayer. The rate of work done by active forces is given by
\begin{equation}
\dot{W}_a=\sum_{i=1}^{N(t)}f_c\dot{a}_i\cos{(\theta_i)}/h\;,
\end{equation}
where $f_c=\mu v_0$ is the crawling force on each cell, $\theta_i$ is the angle between cell polarity and motion for the $i^{th}$ cell.
To derive the equation governing changes in cell area, we use the {\it Onsager's variational principle}~\cite{onsager1931,doi1988,doi2011} adapted for active systems~\cite{banerjee2016,zhang2020,wang2021}, which states that irreversible processes follow the dynamic path that minimizes the Rayleighian $\mathcal{R}$, 
given by the sum of the rate of energy dissipated ($\mathcal{D}$), rate of change in free energy of the system ($\dot{E}_\text{mech}$) and the rate of work done by active forces ($\dot{W}_a$):
\begin{equation}
\mathcal{R}=\mathcal{D} + \dot{E}_\text{mech} - \dot{W}_a\;.
\end{equation}
The equation of motion for cell area follows from minimizing $\mathcal{R}$ with respect to $\dot{a}_i$:
\begin{equation}
\mu \dot{a}_i=\mu h v_0\cos{(\theta_i)}-K(a_i-a_0)h^2\;.
\end{equation}
Now, we define $A(t)=\sum_{i=1}^{N(t)} a_i(t)$ as the total spread area of the monolayer, and $A_0(t)=a_0 N(t)$ as the total target area. Net self-propulsion speed is given by {\color{black}$v_0\sum_{i=1}^{N(t)} \cos{(\theta_i)}\approx v_0 N(t)=v_0 A_0(t)/a_0$, where we made a simple mean-field assumption that cells are polarized in the direction of spreading, $\theta_i\approx 0$, thereby neglecting spatial variations in cell polarity in the spreading monolayer.} Defining $V_0=hv_0/a_0$, we get the following simple equation describing the dynamics of the spread area of the monolayer:
\begin{equation}
\dot{A}=V_0 A_0(t)-k(A(t)-A_0(t))\;,
\end{equation}
where $k = K h^2/\mu$ is the stress relaxation rate of the cell monolayer. Both $V_0$ and $\mu$ are functions of the substrate stiffness $E$. In this mean-field model, traction stress is simply given by $-K(A-A_0)$. The above equation is supplemented by the equation for cell insertion in the monolayer at a rate $g$:
\begin{equation}
\dot{A}_0=gA_0\;.
\end{equation}

The time-dependent solution for monolayer area is given by:
\begin{equation}
A(t) = A(0)\left[\left(1 - \frac{V_0 + k}{g + k}\right) e^{-kt} + \frac{V_0 + k}{g + k} e^{gt}\right]\;,
\end{equation}
which can be approximated at long times as $A(t)\approx A(0)e^{gt} (V_0+k)/(g+k)$. Since both cell crawling speed $v_0$ and friction $\mu$ increases linearly with substrate stiffness $E$, we expect $V_0\propto E$ and $k\propto 1/E$. Therefore, on soft substrates $k\gg V_0$, $A(t) \approx A(0)e^{gt}$, such that monolayer spreading is solely governed by cell addition rate $g$ and is independent of tissue stiffness. On stiff substrates and for stiff tissues, $V_0\gg k$, such that the spread area $A(t)\approx A(0)e^{gt} V_0/g$ is regulated by cell crawling speed independent of tissue mechanical properties. By contrast, soft tissues spread to a larger area on stiff substrates as compared to stiff tissues. 
\begin{figure}
	\centering
	\resizebox{\columnwidth}{!}{
		\includegraphics{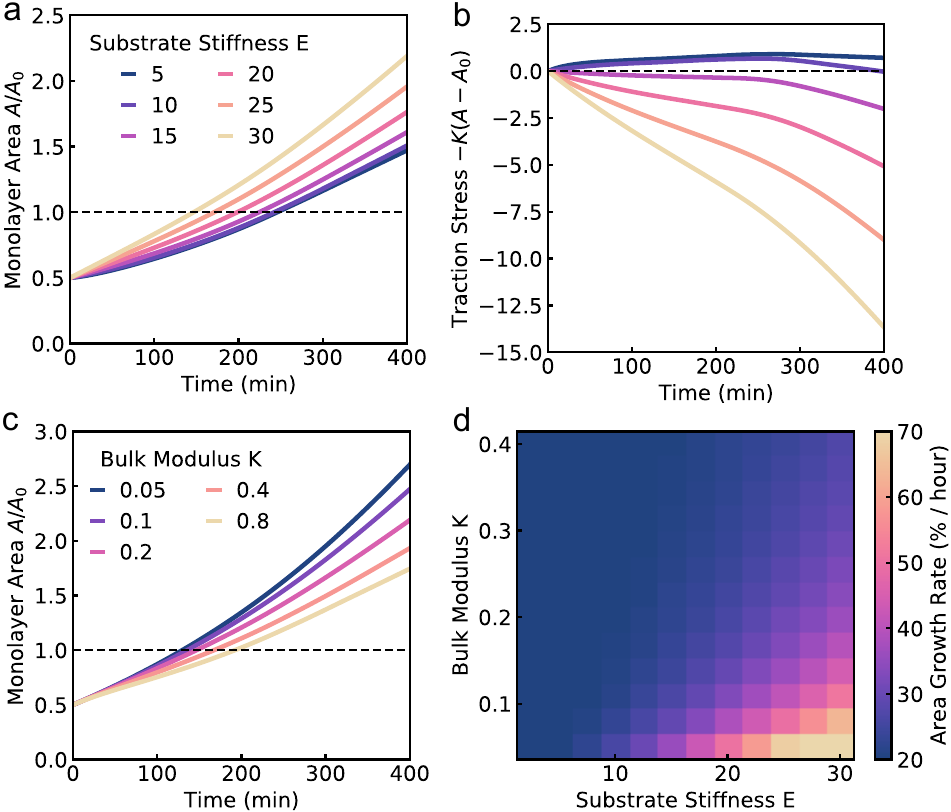}
	}
	\caption{Predictions of the continuum model for spreading monolayers. (a) Model results for the dynamics of monolayer area, and (b) total radial traction stress over time for different values of substrate stiffness. (c) Monolayer area over time for different values of tissue area elastic modulus. (d) Heatmap showing the long-time spreading rates of the aggregate for varying substrate stiffness and tissue elastic modulus.}
	\label{fig:4}     
\end{figure}

If instead, the monolayer reference area expands at a constant rate due to a constant flux of cells from the aggregate, as reported experimentally~\cite{yousafzai2020tissue}, we have
\begin{equation}
\dot{A}_0=g A(0)\;.
\end{equation}
This results in the following time-dependent solution for the monolayer area:
\begin{equation}\label{eq:A}
A(t) = A(0) (V_0 + k)\left(\frac{k -g}{k^2} (1 - e^{-kt}) + \frac{1}{V_0 + k}e^{-kt} + \frac{g}{k}t\right)\;.
\end{equation}
Under this model, we may approximate the monolayer area at long time as $A(t) = A(0) \frac{g(V_0 + k)}{k}t$. Thus, at long times the monolayer spread rate increases with the growth rate or crawl speed, but decreases with tissue stiffness.
Setting the cell crawl speed and friction to increase with substrate stiffness, $V_0 \propto E$ and $k \propto 1/ E$, and fitting the remaining parameters to vertex model simulations, we are able to recapitulate the trends in spread rate and traction stress as substrate stiffness is varied (Fig.~\ref{fig:4}a-b). On soft substrates, spreading is driven by pressure from cell influx. Growth of newly added cells causes an increases in the rest area of the monolayer, $A_0$, resulting in an exponential growth in monolayer area and positive traction stress. As substrate stiffness increases, cell crawling dominates tissue spreading with monolayer area increasing linearly, outpacing the influx of cells and generating negative traction stresses. At long times, we observe a constant rate of area increase that increases with substrate stiffness, consistent with experimental data~\cite{yousafzai2020tissue}.

\begin{figure*}[t]
\includegraphics[width=2\columnwidth]{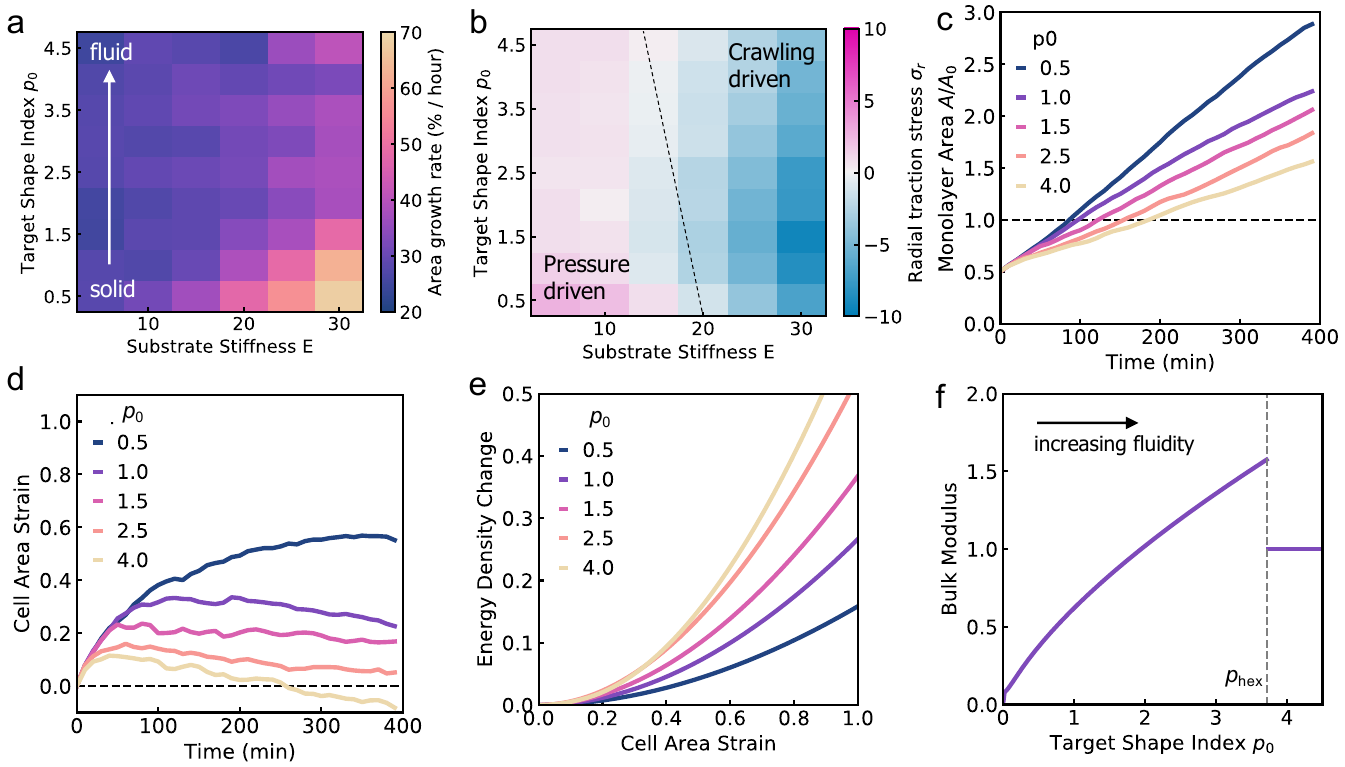}
\caption{{\color{black}Solid-like tissues spread faster than fluid-like tissues.} (a) Area growth rate, and (b) the total radial traction stress for varying substrate stiffness and target shape index $p_0$. (c) Monolayer area over time for varying target shape index. (d) Mean cell area strain over time for varying target shape index. (e) Change in mechanical energy of the tissue per unit area versus cell area strain for varying cell shape indices. (f) Bulk modulus of the tissue versus the target shape index $p_0$. The dashed vertical line indicates the perimeter of a hexagon, $p_\text{hex}$, with unit area.}
	\label{fig:5}     
\end{figure*}

\section{Tissue fluidity impedes collective cell spreading independent of substrate rigidity}
\label{sec:4}

The continuum model for monolayer spreading predicts a simple relationship between spreading rate, substrate rigidity and tissue elasticity. This is evident from Eq.~\eqref{eq:A}, which leads to the relation $A(t) \approx A(0) g(1+V_0/k)t$ as $t\gg k^{-1}$. As the tissue stress relaxation rate $k$ (or equivalently tissue bulk modulus $K$) is reduced or if substrate rigidity $E$ is increased, the model predicts a faster rate of monolayer spreading (Fig.~\ref{fig:4}c). However, this spreading rate also depends on the speed of cell crawling, which in turn is regulated by substrate rigidity. If the crawling speeds are low, as on soft substrates, then the monolayer spreading is driven by cell influx and spreading rate is not affected by the tissue elasticity (Fig.~\ref{fig:4}d). By contrast, if the tissue has a higher relaxation rate $k$, then the spreading rate is less sensitive to tissue and substrate mechanical properties, as the effect of increase in crawl speed is counteracted by the higher tissue bulk modulus (Fig.~\ref{fig:4}d).

To validate the predictions of our continuum model, we sought to investigate how the mechanical properties of the cells influence aggregate spreading, on soft to stiff substrates, using our active vertex model. We vary the target shape index of the cells, $p_0$, to control tissue material properties. When $p_0$ is low, cells are under high tension and the tissue behaves like a jammed solid. When $p_0$ is above a critical value, $p_0 > 3.81$ (in the absence of activity), cells are under no tension and the tissue is in a fluid state, {\color{black} in which the tissue has zero shear modulus}~\cite{bi2015density}. In this state, the cells are able to flow around each other and rearrange with no energy cost. {\color{black}We note that the critical target shape index for rigidity transition depends on the procedure used to generate polygonal tiling. In recent work~\cite{merkel2019,tong2022} it has been shown that the critical $p_0$ can be larger than 3.81 in the presence of vertices with coordination number greater than or equal to four and with cells having five or less neighbors.} When we vary both the target shape index and the substrate stiffness, we find that on soft substrates, changing $p_0$ has little impact on monolayer spreading rate (Fig.~\ref{fig:5}a). On soft substrates, tissue spreading is pressure-driven (Fig.~\ref{fig:5}b), with the aggregate adding the same area of cells per unit time. Thus monolayer spreading can only be slowed down by modulating the friction with the substrate. Changing the fluidity or tension of the tissue has little effect on the isotropic bulk pressure driving monolayer spreading. However, when spreading occurs on stiff substrates, we find that cells with a lower $p_0$, and thus higher tension, spread faster, with the monolayer spreading twice as fast at $p_0 = 0.5$ compared to $p_0 = 4.5$ (Fig.~\ref{fig:5}c). Moreover, the mechanosensitivity of monolayer spreading reduces as $p_0$ increases. At high $p_0$, changes in substrate stiffness have less effect on the rate of spreading than for low $p_0$.

Why do fluid tissues spread slower than solid tissues (Fig.~\ref{fig:5}c)? While it is expected that cells in a fluidized tissue move faster as they can rearrange more easily~\cite{tetley2019tissue}, monolayer spreading is driven by a radially outward stress arising from active cell crawling or isotropic pressure from cell influx, where tissue shear modulus does not play a role. Fluid tissues have a zero shear modulus~\cite{bi2015density}, but can still maintain a finite bulk modulus that can resist isotropic expansion of the tissue. To determine the relationship between tissue fluidity and bulk modulus, we computed cell area strain during spreading for a range of $p_0$ values. We find that cells are much more strained at low $p_0$, with cells area increasing by around 60\% for $p_0 = 0.5$, compared to cells area maintained at $p_0 = 4.5$ (Fig.~\ref{fig:5}d).  When we compute the energy cost per unit area for such bulk deformations in the vertex model with perfectly hexagonal cells, cells with a lower $p_0$ require much less energy for deformation than cells with a higher $p_0$, despite low $p_0$ cells being under higher tension (Fig.~\ref{fig:5}e). These data suggest a counterintuitive result that tissue bulk modulus increases with increasing $p_0$ or increasing tissue fluidity. Thus, on stiff gels where we have high forces generated by active cell crawling, decreasing tissue fluidity also reduces the bulk modulus, allows for a faster rate of monolayer spreading.

To quantify the relationship between tissue bulk modulus and target shape index, we calculate the bulk modulus $K$ as the second derivative of energy density with respect to the area strain $\varepsilon$, $K=(1/a^*) \partial^2 \tilde{E} / \partial \varepsilon^2$, where $a^*$ is the area of the cell at equilibrium, and $\tilde{E}$ is given by Eq.~\eqref{eq:energy}. We find that the bulk modulus increases monotonically with $p_0$ for $p_0 < p_\text{hex} \approx 3.722$, where $p_\text{hex}$ is the perimeter of a hexagon of unit area (Fig.~5f). Interestingly, there is a discontinuity in bulk modulus for $p_0 > p_\text{hex}$, beyond which the bulk modulus stays constant. To gain a mechanistic understanding of these results, we note that the cell area and perimeter can be written as $a = a^* (1 + \varepsilon)$ and $p = p^* (1 + \varepsilon)^{1/2}$. {\color{black}As previously shown by Staple et al~\cite{staple2010mechanics},} taking the second derivative of $E_\text{mech}$ with respect to $\varepsilon$  and evaluating the energy at $\varepsilon = 0$ one obtains $K = a^* + \Gamma p^* p_0 / 4 a^*$. For $p_0 < p_\text{hex}$, $a^* \propto p_0$ and $p^*\propto{a^*}^{1/2}$. The second term in $K$ thus scales like ${a^*}^{1/2}$, resulting in an increase in bulk modulus with increasing cell area, which in turn increases with $p_0$ (Fig.~\ref{fig:5}f). Since low $p_0$ cells are under higher tension they are also smaller in size, resulting in a lower bulk modulus (Fig.~5f). Having a low $p_0$, high tension cell is much like how two springs in series are softer than a single spring: while the energy cost per unit area is similar, the deformation is spread between more cells and so we have a lower overall energy cost. However, for $p_0 > p_\text{hex}$, cells adapt their perimeters to the target perimeter. As a result, the perimeter term doesn't contribute to energy in Eq.~\eqref{eq:energy}, resulting in a discontinuity in the bulk modulus. Taken together, our theory and simulations reveal the interdependence between cell shape, tissue elasticity and substrate rigidity in the spreading of cell monolayers on compliant substrates.

\section{Discussion}
In this study, we investigated how tissue mechanics and substrate stiffness, and their interplay, regulate the dynamics of a spreading cellular aggregate, using a combination of computational simulations and mathematical modeling. In particular, we developed an active vertex model to simulate the spreading of a cellular monolayer emanating from a three-dimensional aggregate, which is undergoing active growth as well as driven forward by cell crawling. Our simulations reveal two distinct modes of cell monolayer spreading depending on substrate rigidity. On soft substrates, cell monolayer spreading is pressure-driven, exhibiting radially outward traction stresses that originate from the influx of cells into the monolayer from the aggregate. By contrast, on stiff substrates, cell crawling forces drive monolayer expansion, generating inward traction forces localized to the periphery of the cell monolayer. Despite the different mechanisms for spreading, our simulation reveals comparable spreading rates on substrates of varying rigidity, consistent with experimental data~\cite{yousafzai2020tissue}. {\color{black}The rigidity-dependent transition from pressure-based to traction-based spreading arises in the model because of the cell-substrate coupling that increases cell polarisation, crawl speed, and traction forces with increased substrate rigidity~\cite{discher2005tissue, zemel2010, trichet2012, oakes2014, yousafzai2020tissue}.}

The modes of collective cell motion and the rate of spreading is not only dependent on substrate rigidity but can also be tuned by varying the rate of cell influx from the aggregate and the speed of cell crawling. When the cell crawling speed is high relative to the cell influx rate, crawling-driven spreading dominates, with inward traction forces localized to the tissue periphery. By contrast, when crawling speeds are lower compared to cell influx rate, pressure-driven spreading drives monolayer expansion with outward traction forces distributed throughout the monolayer. However, we find that the cell addition rate is the main regulator of spreading rates, with crawling speeds only able to increase spread rates by 50\%, suggesting that the influx of new cells is the limiting factor for spreading.

To further understand the cellular mechanisms controlling spreading rates, we develop a simple continuum model of the spreading tissue as an elastic medium with active growth and cell crawling. We identify the tissue bulk modulus as another important mechanical property governing the dynamics of spreading. When the cells are very stiff with a high bulk elastic modulus, our model predicts little variation of spread rates with a change in substrate stiffness or crawl speed, with the rate of cell addition governing the spreading rates. However, when the cells are soft, traction forces generated by crawling cells may produce larger strains on cells. This results in spreading rates that are very sensitive to the mechanical stiffness of the underlying substrate. To tests this prediction, we used our active vertex model to simulate monolayer spreading by varying cell stiffness through changes in the target shape index $p_0$. When the target shape index is low, cells are under high tension but have a reduced bulk modulus. By contrast, cells with a higher target shape index result in a fluid tissue that has a high bulk modulus. As a result, cell monolayers with a lower target shape index (and consequently lower bulk modulus) spread faster on stiff substrates than on soft substrates.

Overall, these results capture previously reported data on the spreading of cellular aggregates with increased spreading rates on stiff substrates, and also explain the spatiotemporal patterns in traction stresses on soft substrates driven by active pressure, and on stiff substrates driven by active cell crawling. Moreover, our theory and simulations provide new predictions on the role of tissue mechanics on cell spreading, with fluid tissues being less sensitive {\color{black}to substrate rigidity and spread slower than solid tissues.}\\
 
{\bf Acknowledgements}. SB acknowledges funding from HFSP RGY0073/2018 and NIH R35 GM143042. MPM acknowledges funding from ARO MURI W911NF-14-1-0403, NIH RO1 GM126256, NIH U54 CA209992 and HFSP RGY0073/2018.\\


%

\end{document}